\documentclass[showpacs,floatfix,nofootinbib]{revtex4}

\usepackage{amsmath,amssymb}

\newcommand{\be}{\begin{equation}}
\newcommand{\ee}{\end{equation}}
\newcommand{\eq}[1]{Eq.~(\ref{#1})}
\newcommand{\eqs}[1]{Eqs.~(\ref{#1})}

\newcommand{\RR}{\mathbb{R}}
\newcommand{\CC}{\mathbb{C}}
\newcommand{\rr}{\boldsymbol{r}}
\newcommand{\s}{\boldsymbol{s}}
\newcommand{\HH}{\boldsymbol{H}}
\newcommand{\BB}{\boldsymbol{B}}
\newcommand{\JJ}{\boldsymbol{J}}
\newcommand{\A}{\boldsymbol{A}}

\newcommand{\ket}{|\psi\rangle}
\newcommand{\bra}{\langle\psi|}

\newcommand{\ketr}{|\psi(\rr)\rangle}

\newcommand{\ketp}{|\phi\rangle}
\newcommand{\brap}{\langle\phi|}

\begin{document}

\title{Singular solutions to the Seiberg-Witten and Freund \\equations on flat space
       from an iterative method }

\author{Ricardo A. Mosna}
\email{mosna@ime.unicamp.br}
\affiliation{
Instituto de Matem\'atica, Estat\'\i stica e Computa\c{c}\~ao Cient\'\i fica,
Universidade Estadual de Campinas, C.P. 6065, 13083-859,
Campinas, SP, Brazil.}

\date{\today}


\begin{abstract}
Although it is well known that the Seiberg-Witten equations do not admit
nontrivial $L^2$ solutions in flat space, singular solutions to them have been
previously exhibited --- either in $\mathbb{R}^3$ or in the dimensionally
reduced spaces $\mathbb{R}^2$ and $\mathbb{R}^1$ --- which have physical
interest. In this work, we employ an extension of the Hopf fibration to
obtain an iterative procedure to generate particular singular solutions to the
Seiberg-Witten and Freund equations on flat space. Examples of solutions
obtained by such method are presented and briefly discussed.
\end{abstract}

\maketitle

\section{Introduction}
\label{sec intro}

Given a physical system defined on a configuration space $M$,
there are various instances where it is useful to employ
(extensions of) fibrations $P\rightarrow M$ to lift the
corresponding equations of motion from $M$ to $P$.
For instance, the natural extension of the Hopf fibration
$S^3\rightarrow S^2$ to $\RR^4\rightarrow\RR^3$ (defining the
so-called Kustaanheimo-Stiefel transformation \cite{ks,kibler})
can be used to map the Kepler problem in $\RR^3$ to a
harmonic oscillator problem in $\RR^4$.
This construction has been recurrently employed to regularize and
calculate orbits of celestial objects, besides giving rise to various applications
in atomic physics (see, e.g., \cite{Bartsch} and references therein).
In this work, we apply this idea to the case when $M$, instead of representing the
configuration space of a particle, is the target space of a given field theory.
Specifically, we show that by lifting the equations of magnetostatics
(in the sense above), it is possible to obtain the Seiberg-Witten equations (SWE)
on $\RR^3$ provided that a certain constraint is imposed on the resulting fields.
Moreover, we show that such constraint naturally gives rise to an iterative method
to generate particular solutions to the SWE and Freund equations on $\RR^3$ and
its dimensionally reduced spaces.

It should be kept in mind that the SWE do not admit nontrivial $L^2$ solutions
in flat space \cite{Witten94} (the same is not true for the Freund equations
\cite{AMN}). However, singular solutions to the SWE in flat space
do exist \cite{2d,AMN}, with physical interest.
Another point to be emphasized is that the lifting procedure considered here
(Section~\ref{sec main}) is not new since it is implicit in the pioneer work
of Loss and Yau on zero modes of the three-dimensional Dirac operator \cite{Loss}
(it is also known that, by applying the Kustaanheimo-Stiefel transformation
to the vector potential coupled to a Dirac spinor, one recovers the {\em ansatz}
of Loss and Yau \cite{AT}).
In fact, our formulae for the relevant Abelian potential $A_k$
and wavefunction $\ket$  are the same as those of \cite{Loss}, where $A_k$
and $\ket$ are given in terms of a generating vector field, but with the
differences that, in our case, such generating vector field satisfies
a certain constraint and that, due to the singular nature of the present
problem, we do not demand that the associated fields be square integrable.

It is precisely such constraint that gives rise, in our approach,
to the aforementioned iterative method to the SWE and Freund equations
on $\RR^3$. This is considered in Section~\ref{subsec swe}, where we
also show that application of such method recovers some known
solutions to the SWE and Freund equations and yields, to
the best of our knowledge, previously unnoticed solutions to the SWE.
In particular, we obtain an axisymmetric singular solution to the SWE on $\RR^3$.
We conclude by presenting some final remarks in Section~\ref{sec conc}.

\section{Lifting the magnetostatics equations}
\label{sec main}

We start from the equations of magnetostatics,
\begin{subequations}
\label{magnetostatics}
\begin{align}
\nabla\cdot\HH  &  =0,
\label{magnetostatics a}\\
\nabla\times\HH  &  =\JJ,
\label{magnetostatics b}
\end{align}
\end{subequations}
where $\JJ$
is the steady current associated with the magnetic field $\HH$ (we use
Heaviside-Lorentz units with c=1).
Let $\ket$ be a two-component spinor such that
\begin{equation}
H^k=\bra\sigma^k\ket,\quad k=1,2,3,
\label{Hk}
\end{equation}
where $\sigma^1, \sigma^2, \sigma^3$ are the Pauli matrices.
For each $\rr\in\mathbb{R}^{3}$, $\HH(\rr)$ can be formally regarded as the
``polarization vector'' or ``spin density'' associated with $\ketr\in\mathbb{C}^{2}$,
as in quantum mechanics textbooks \cite{Merzbacher}.
The general solution of \eq{Hk} for $\ket$ in terms of $\HH$ is given by%
\footnote{In spherical coordinates, \eq{psi} assumes the familiar form
$\ket=e^{-i\chi}\sqrt{H}
\begin{pmatrix}
\cos\frac{\alpha}{2}\\
\sin\frac{\alpha}{2}\,e^{i\beta}
\end{pmatrix}$,
where
$\HH=H (  \sin\alpha\cos\beta, \sin\alpha\sin\beta,\cos\alpha )  $.}
\begin{equation}
\ket=e^{-i\chi}\frac{1}{\sqrt{2(H+H^{3})}}
\left(
\begin{array}[c]{c}
H+H^{3}\\
H^{1}+iH^{2}
\end{array}
\right),
\label{psi}
\end{equation}
where$\ H=\lVert \HH\rVert $ and $e^{-i\chi}$ is an arbitrary phase factor.%
\footnote{It is interesting to note that this is a (trivial) application of
what has been termed the inversion theorem \cite{Crawford}, an useful
result (especially in four dimensions \cite{Vaz,quato}) when one wants to reconstruct a
given spinor, apart from arbitrary phases, from its bilinear covariants.}

Before transferring the dynamics (\eqs{magnetostatics}) from $\HH$ to $\ket$,
we briefly consider the geometry underlying \eqs{Hk} and (\ref{psi}).
Let $S^n_a$ denote the $n$-sphere of radius $a$ in $\RR^{n+1}$,
and consider the map $\pi_a:S^3_a\to S^2_{a^2}$ taking a two-component spinor
$\ketp\in S^3_a\subset \CC^2$ into the vector $\s\in S^2_{a^2}$
with components $s^k=\brap\sigma^k\ketp$.%
\footnote{Here $\ketp=
\begin{pmatrix}
z_1 \\
z_2
\end{pmatrix}
\in S^3_a\subset \CC^2$ means that
$\lvert z_1 \rvert^2 + \lvert z_2 \rvert^2 = a^2$.}
This defines a principal fiber bundle $U(1)\cdots S^3_a\to S^2_{a^2}$
which is essentially the first Hopf bundle (where one usually takes $a=1$) \cite{Hopf}.
More generally, one can drop the requirement that $\ketp$ belongs
to a sphere of fixed radius and consider the map $\pi:\RR^4\to \RR^3$
taking $\ketp\in\CC^2\cong\RR^4$ into $s^k=\brap\sigma^k\ketp$.
In this way, $\pi$ is a natural extension of the Hopf map $S^3\to S^2$
to $\RR^4\to \RR^3$, in which each sphere of radius $a>0$ in $\RR^4$
is mapped into the sphere $S^2_{a^2}\subset\RR^3$,
and the origin of $\RR^4$ is mapped into the origin of $\RR^3$.
Such map defines the so-called Kustaanheimo-Stiefel transformation \cite{ks,kibler}
(this no longer gives rise to a principal fiber bundle, since the fiber over the
origin is just a point).
Note that, in our case, \eq{Hk} defines a Kustaanheimo-Stiefel transformation on the
corresponding {\em target spaces}, relating, for each $\rr$, the vector
$\HH(\rr)\in\RR^3$ to the spinor $\ketr\in\CC^2\cong\RR^4$.
We also note that \eq{psi} yields, for each fixed $\rr$,
a local section of the bundle $U(1)\cdots S^3_H\to S^2_{H^2}$
over $S^2_{H^2}\setminus\{\textrm{south pole}\}$. A related
local section over $S^2_{H^2}\setminus\{\textrm{north pole}\}$ can
be similarly obtained.

Going back to \eq{psi}, it is easy to see that the density matrix associated with
$\ket$ is given by
\begin{equation}
\ket\bra=\frac{1}{2}\left(  H\openone +H^{k}\sigma_{k}\right), \label{density matrix}
\end{equation}
where $\openone$ is the identity $2\times 2$ matrix
(notice that we are working with Cartesian coordinates in Euclidean flat space,
so that indices can be freely raised and lowered).
For what follows, it is useful to define the following matrix-valued functions:
\begin{subequations}
\label{matrix field}
\begin{align}
\mathsf{H}(\rr)  &  =H^{k}(\rr)\sigma_{k},
\label{matrix field a}\\
\mathsf{J}(\rr)  &  =J^{k}(\rr)\sigma_{k}.
\label{matrix field b}
\end{align}
\end{subequations}
Then, it is easily seen that \eqs{magnetostatics} can be equivalently
written, in terms of $\mathsf{H}$ and $\mathsf{J}$, as
\begin{equation}
\partial\mathsf{H}=i\mathsf{J},
\label{matrix magnetostatics}
\end{equation}
where $\partial=\sigma^k\partial_k$ (this follows at once from the
relationship $\sigma^i \sigma^j =\delta^{ij}\openone+i\epsilon^{ijk}\sigma_k$
satisfied by the Pauli matrices, where $\epsilon^{ijk}$ is the totally
antisymmetric symbol with $\epsilon^{123}=1$).

We now transfer the dynamics defined by \eqs{magnetostatics}
from $\HH$ to $\ket$. From
\eq{density matrix}:
\[
\mathsf{H}=2\ket\bra-H\openone,
\]
which leads, upon substitution into \eq{matrix magnetostatics}, to
\[
\sigma^{k}\left[  \partial_{k}\ket\bra+\ket
\partial_{k}\bra-\frac{1}{2}\partial_{k}H-\frac{i}{2}J_{k}\right]
=0.
\]
Our aim is to obtain a differential equation governing the dynamics of $\ket$.
To that end, we right-multiply the above equation by
$\ket$ and use the fact that $\langle\psi\ket=H$. This yields
\begin{equation}
\sigma^{k}\left[  \partial_{k}+\frac{1}{2H}\partial_{k}H-\frac{1}{H}
\bra\partial_{k}\ket-\frac{i}{2H}J_{k}\right]  \ket=0.
\label{aux}
\end{equation}
The term $\bra\partial_{k}\ket$ can be computed by a
straightforward calculation; it follows from \eq{Hk} that
\[
\bra\partial_{k}\ket=\frac{1}{2}\partial_{k}H+\frac
{i}{2(H+H^{3})}\left(  H^{1}\partial_{k}H^{2}-H^{2}\partial_{k}H^{1}\right)
-iH\partial_{k}\chi.
\]
Upon substitution into \eq{aux}, this leads to
\[
\sigma^{k}\left[  \partial_{k}+
i \left(
\partial_{k}\chi-
\frac{1}{2H(H+H^{3})} \left(  H^{1}\partial_{k}H^{2}-H^{2}\partial_{k}H^{1}\right)
-\frac{1}{2H}J_{k}
\right)
\right]  \ket=0.
\]
Defining
\begin{equation}
A_{k}:=-\frac{1}{2H(H+H^{3})}
\left(  H^{1}\partial_{k}H^{2}-H^{2}\partial_{k}H^{1}\right)-
\frac{1}{2H}J_{k},
\label{Ak0}
\end{equation}
which can be fully expressed in terms of $\HH$ (through  \eq{magnetostatics b}) as
\begin{equation}
A_{k}=-\frac{1}{2H(H+H^{3})}
\left(  H^{1}\partial_{k}H^{2}-H^{2}\partial_{k}H^{1}\right)-
\frac{1}{2H}(\nabla\times\HH)_k,
\label{Ak}
\end{equation}
we finally get
\begin{equation}
i \sigma^{k}\left(  \partial_{k} + i A_{k} + i \partial_{k}\chi\right)  |\psi \rangle=0.
\label{Weyl eq}
\end{equation}
Therefore, $\ket$ satisfies the Weyl equation\footnote{That is, the massless Dirac equation
for (two-component) spinors representing states of definite chirality.} with the Abelian
potential $A_{k}$. Note that $\chi$ enters \eqs{psi} and (\ref{Ak}) simply as a gauge parameter.

It is interesting to note that \eq{magnetostatics b} enters the derivation
above merely as a bookkeeping device. In fact, \eq{Weyl eq} for $\ket$ follows
as long as $\HH$ satisfies \eq{magnetostatics a}, regardless of any interpretation
of the right-hand side of \eq{magnetostatics b} as an external current.
In any case, it should be noted that \eq{magnetostatics b} does affect the
form of \eq{Weyl eq} through $A_k$.

An important observation for what follows is that the field strength $B_k$ associated with $A_{k}$,
\begin{equation}
\BB:=\nabla\times\A,
\label{B}
\end{equation}
does not have to bear any relation to the magnetic field $\HH$ we started with.

\section{Seiberg-Witten and Freund equations}
\label{subsec swe}

Let us summarize what has been done above. We started from the equations of
magnetostatics, expressed the magnetic field $\HH$ in terms of the
associated spinor field $\ket$, and then lifted the dynamics from
$\HH$ to $\ket$. As a result, the following set of
equations (\eqs{Hk}, (\ref{Weyl eq}), and (\ref{B})) was obtained:
\begin{subequations}
\label{main equations}
\begin{align}
\bra\sigma^k \ket &= H^k,
\label{main equations a}\\
i \sigma^{k}(\partial_{k} + i A_{k})\ket &=0,
\label{main equations b}\\
\epsilon^{ijk}\partial_{i}A_{j} &=B^k,
\label{main equations c}
\end{align}
\end{subequations}
where we chose to suppress the terms associated with the gauge parameter $\chi$.
The SWE and Freund equations in three dimensions have been discussed in detail
in \cite{AMN} (see especially its equations (3.5) and (3.6)) from where
we note a remarkable similarity with \eqs{main equations}. More precisely:

\begin{enumerate}
\item Eqs.~(\ref{main equations}) are the Seiberg-Witten equations on $\RR^3$
\emph{provided that} $H_{k}=+B_{k}$;
\item Eqs.~(\ref{main equations}) are the Freund equations on $\RR^3$
\emph{provided that} $H_{k}=-B_{k}$.
\end{enumerate}

Therefore, the constraint
\begin{equation}
\HH=\pm\BB
\label{H=B}
\end{equation}
yields a natural \emph{ansatz} for obtaining solutions to the Seiberg-Witten
and Freund equations for $\A$, $\BB$, and $\ket$ on $\mathbb{R}^{3}$.
Using \eq{Ak}, this amounts to solving
\begin{equation}
\HH=\pm\nabla\times\left(
-\frac{1}{2H(H+H^{3})}\left(  H^{1}\nabla H^{2}-H^{2}\nabla H^{1}\right)-
\frac{1}{2H}\nabla\times\HH
\right)
\label{eq for H}
\end{equation}
for $\HH$. This equation has been recently studied from a group-theoretical
perspective in \cite{AT2} to examine the Lie symmetries of the SWE and
Freund equations on $\RR^3$.

It is interesting to note that, given a solution of \eq{eq for H}, with the $+$ or $-$
sign, respectively, one immediately obtains $\ket$, $A_k$, and $B_k$ from \eqs{psi},
(\ref{Ak}), and~(\ref{H=B}):
\begin{subequations}
\label{A and psi}
\begin{align}
\ket &  =\frac{1}{\sqrt{2(H+H^{3})}}
\left(
\begin{array}[c]{c}
H+H^{3}\\
H^{1}+iH^{2}
\end{array}
\right),   \label{A and psi b} \\
\A  &  =  -\frac{1}{2H(H+H^{3})}\left( H^{1}\nabla H^{2}-H^{2}\nabla H^{1} \right)  -
\frac{1}{2H}\nabla\times\HH, \label{A and psi a} \\
\BB &  =  \pm\HH. \label{A and psi z}
\end{align}
\end{subequations}
As noted in the Introduction, the above expressions for $\ket$ and $\A$ in terms of a
generating vector field (which is given, in this case, by $\HH$) were first obtained
in the study of zero modes of the massless Dirac operator in \cite{Loss}.

\subsection{An iterative procedure}
\label{sec ite}

We now show how \eq{eq for H} can serve as a basis for an iterative procedure
for obtaining $\HH$, and thus $\ket$, $\A$ and $\BB$ satisfying
the SWE or Freund equations on flat space. The procedure goes as follows.
Choose an initial guess $\HH_{\!(0)}$ for $\HH$;
substitute $\HH_{\!(0)}$ into the right-hand side of \eq{eq for H} and
consider the result as a second estimate $\HH_{\!(1)}$ for $\HH$;
then substitute $\HH_{\!(1)}$ into the right-hand side of \eq{eq for H},
and so on.
If the sequence $\HH_{\!(k)}$ converges, its limit is a solution to \eq{eq for H}.
It is important to note that this procedure does fail in most cases, either by
computational or mathematical difficulties (we come back to this point in
Section~\ref{sec conc}). Nevertheless, when it succeeds, we end up with a
solution to the Seiberg-Witten or Freund equations.
In the remainder of this section, we show representative results of a limited experiment
in algebraic computation, performed with the software Mathematica,
implementing such iterative procedure.

\medskip

{\bf Example 1. }
Starting with $\HH_{\!(0)}=\pm(x,y,z)$, we obtain the solution
\begin{subequations}
\begin{align}
\BB  & = \mp\frac{1}{2r^{3}} \; (x,y,z), \\
\A   & = \frac{1}{2r(r\pm z)} \; \left( y,-x,0 \right), \\
\ket & = \frac{1}{2r\sqrt{r(r\pm z)}}
\begin{pmatrix}
r\pm z \\
\pm (x+iy)
\end{pmatrix}
\end{align}
\end{subequations}
to the Freund equations, where $r=\sqrt{x^2+y^2+z^2}$.
A monopole solution of this kind was first obtained in \cite{Freund}
(see also~\cite{AMN}, where the authors discuss in detail
how the Freund equations are related to the SWE on $\RR^3$).

\medskip

{\bf Example 2. }
Starting with $\HH_{\!(0)}=\pm(\sinh\kappa y,0,0)$,
we obtain the solution%
\footnote{In order to avoid dealing with the absolute value function
in the algebraic computation procedure of Examples 2 and 3,
it is useful to first restrict attention to the domain given
by $x>0$, $y>0$, and $z>0$, and later extend the obtained solution
to any nonzero $x$, $y$, and~$z$.}
\begin{subequations}
\label{swe1d}
\begin{align}
\BB  & = \mp\frac{\kappa^{2}}{\sinh^2\kappa y} \; \mathbf{e}_x, \label{swe1d b} \\
\A   & = \pm\kappa \coth\kappa y \; \mathbf{e}_z,               \label{swe1d a} \\
\ket & = \frac{\kappa}{\sqrt{2}\sinh\kappa y}
\begin{pmatrix}
1 \\
\mp 1
\end{pmatrix}
\end{align}
\end{subequations}
to the Seiberg-Witten equations. This solution is essentially the same
as the effectively one-dimensional solution to the SWE found in \cite{2d}.
On the other hand, if we start with $\HH_{\!(0)}=\pm(\cosh\kappa y,0,0)$,
we obtain the solution
\begin{align*}
\BB  & = \pm\frac{\kappa^{2}}{\cosh^2\kappa y} \; \mathbf{e}_x, \\
\A   & = \pm\kappa \tanh\kappa y \; \mathbf{e}_z, \\
\ket & = \frac{\kappa}{\sqrt{2}\cosh\kappa y}
\begin{pmatrix}
1 \\
\mp 1
\end{pmatrix}
\end{align*}
to the Freund equations. We note that similar expressions were also obtained in~\cite{2d}
through analytic continuation of the aforementioned one-dimensional solution to the SWE.

\medskip

{\bf Example 3. }
Starting with $\HH_{\!(0)}=\pm(xyz,0,0)$, we obtain the solution
\begin{subequations}
\label{swe2d}
\begin{align}
\BB  & = \mp\left(\frac{1}{y^{2}}+\frac{1}{z^{2}},0,0\right), \label{swe2d b} \\
\A   & = \pm\frac{1}{y^2+z^2}\left(0,-\frac{y^2}{z},\frac{z^2}{y}\right), \label{swe2d a} \\
\ket & = \sqrt{\frac{1}{2y^{2}}+\frac{1}{2z^{2}}}
\begin{pmatrix}
1 \\
\mp 1
\end{pmatrix}
\end{align}
\end{subequations}
to the Seiberg-Witten equations.

For solutions of this kind, in which $\BB(\rr)$ is always parallel to some fixed
vector $\mathbf{n}$ and only depends on coordinates $(u,v)$ of a plane orthogonal to
$\mathbf{n}$, the quantity $\omega=\frac{1}{2}\ln B$ is known \cite{2d} to
satisfy the Liouville equation $4\partial_z\partial_{\bar{z}}\omega=e^{2\omega}$,
where $z$ now denotes the complex coordinate $z=u+iv$ and $B=\lVert\BB\rVert$.%
\footnote{It should be clear from the context when $z$ refers to the complex
coordinate $z=u+iv$ or to the Cartesian coordinate in $(x,y,z)$.}
Using the {\em ansatz}
\begin{equation}
\omega=\frac{1}{2}\ln\frac{4 (dg/dz)(d\bar{g}/d\bar{z})}{(1-g\bar{g})^2},
\label{ansatz_NS}
\end{equation}
with $g(z)$ an arbitrary analytic function, the authors of~\cite{2d}
construct a family of effectively two-dimensional solutions to the SWE
with interesting properties. We note, however, that the
above solution~(\ref{swe2d b}) apparently does not belong to such family,
obtained via {\em ansatz}~(\ref{ansatz_NS}).
In any case, we show in the Appendix that the alternative {\em ansatz}%
\footnote{It should be noted that both (\ref{ansatz_NS}) and (\ref{ansatz_alt}) are
particular cases of the well-known general solution of the Liouville equation
\cite{liouville} (see Appendix).}
\begin{equation}
\omega=\frac{1}{2}\ln\frac{(dg/dz)(d\bar{g}/d\bar{z})}{[\Im(g)]^2},
\label{ansatz_alt}
\end{equation}
where $\Im(g)$ denotes the imaginary part of $g$, does yield the above solution.
In fact, as discussed in the Appendix, (\ref{swe2d b}) is the $n=2$ case of
a family of two-dimensional singular solutions generated by the choice
$g(z)=z^n$ in (\ref{ansatz_alt}), with $n=\frac{1}{2},1,\frac{3}{2},\ldots$ .

\medskip

{\bf Example 4. }
Starting with $\HH_{\!(0)}=\pm(y,-x,0)$, we obtain the axisymmetric solution
\begin{subequations}
\label{B_ax}
\begin{align}
\BB  &= \pm\frac{1}{2\rho^{2}}\mathbf{e}_{\phi},                        \label{B_ax b} \\
\A   &= -\frac{1}{2\rho}\mathbf{e}_\phi \pm\frac{1}{2\rho}\mathbf{e}_z,  \label{B_ax a} \\
\ket & = \frac{1}{2\rho}
\begin{pmatrix}
1 \\
\pm ie^{i\phi}
\end{pmatrix}
\end{align}
\end{subequations}
to the Seiberg-Witten equations, where cylindrical coordinates $\rho,\phi,z$,
with $\rho=\sqrt{x^{2}+y^{2}}$ and $\phi=\arctan(y/x)$, were used.%
\footnote{Such $\BB$ is similar (but different in the $\rho$-dependence)
to the magnetic field $\BB\propto\frac{1}{\rho}\mathbf{e}_{\phi}$ produced by
a steady current along the $z$ axis.}
The integral curves of $\A$ are helices of constant $\rho$ going upward (downward)
with respect to the $z$~axis.
We note that the first term $\A_{AB}= -\frac{1}{2\rho}\mathbf{e}_\phi$
of $\A$ is in fact an Aharonov-Bohm potential, with holonomy given by
\be
\int_{\gamma} \A_{AB}(\rr)\cdot d\rr= -\pi,
\label{flux}
\ee
where $\gamma$ is any loop winding once around the $z$~axis.
Note that the Aharonov-Bohm term $\A_{AB}$ is actually {\em implied}
by $\BB$ (through \eq{A and psi a}) even though $\BB$ in \eq{B_ax b} does not receive
any contribution from the curl of $\A_{AB}$ (which is actually zero for $\rho\ne0$).
In this way, the purely azimuthal magnetic field of \eq{B_ax b}, defined away
from the $z$~axis, unavoidably gives rise, through $\A_{AB}$, to the
``additional'' singular magnetic field $-\pi\,\delta(x)\delta(y)\mathbf{e}_z$
along the $z$~axis (which is the same as the magnetic field of an infinitely
long and infinitesimally thin solenoid at the $z$~axis).

It is also interesting to note that, although the two solutions
in \eq{B_ax b} (given, respectively, by its plus and minus signs)
wind in opposite directions with respect to the $xy$ plane,
their associated potentials both wind clockwise, with identical
Aharonov-Bohm terms, the only difference residing in their $z$ components.

\section{Closing remarks}
\label{sec conc}

In all the examples above, the final solution for $\HH$ is obtained in
\emph{exact form} after very few iterations. On the other hand, some experience
with the above computational experiment shows that a generic initial
condition for $\HH$ typically leads to an increasingly complicated
algebraic expression at each iteration, thereby requiring further
investigation on convergence issues related to such method.
A natural question to ask is what are the initial conditions under
which the sequence $\HH_{\!(k)}$ may be guaranteed to converge since,
under such circumstances, the approach presented here could be used
to define classes of solutions to the SWE and Freund equations iteratively.

Finally, we note that the approach presented here suggests
a natural generalization to the four-dimensional case, where
one may try to lift the whole set of (Euclidean) Maxwell equations
to obtain (singular) solutions to the SWE on $\RR^4$.
This is the subject of work in progress.

\acknowledgments

The author is grateful to C. Adam, W. A. Rodrigues Jr., M. A. F. Rosa, A. Saa
and J. Vaz Jr. for helpful discussions, and to the Abdus Salam International
Centre for Theoretical Physics, Trieste, Italy, where this work was
partly done, for hospitality.
This work was supported by FAPESP.

\appendix

\section*{Appendix}
The general solution to the Liouville equation,
\[
4\,\partial_z\partial_{\bar{z}}\omega=e^{2\omega},
\]
was first given by Liouville in \cite{liouville} (where the author considers,
in fact, the corresponding equation for $\lambda=e^{2\omega}$, and with real variables).
It is given by
\begin{equation}
\omega(z,\bar{z})=\frac{1}{2}\ln\frac{4(dg(z)/dz)(dh(\bar{z})/d\bar{z})}{[1-g(z)h(\bar{z})]^2},
\label{gensol}
\end{equation}
where $g(z)$ and $h(\bar{z})$ are arbitrary analytic and anti-analytic functions,
respectively. The {\em ansatz} of \cite{2d}, given by \eq{ansatz_NS}, is recovered
from \eq{gensol} if the natural choice $h(\bar{z})=\overline{g(z)}$ is made.
However, as mentioned in Example~3 of Section~\ref{subsec swe}, the solution given
by $B=1/u^2+1/v^2$ is apparently not recovered by such {\em ansatz} (recall that $B$ is related
to $\omega$ by $B=e^{2\omega}$, where $B=\lVert\BB\rVert$ and now $z$ denotes the complex coordinate
$z=u+iv$).
This motivates the search for an alternative {\em ansatz} for such $B$.
Generalizing the above choice of $h$ in terms of $g$ to
$h(\bar{z})=\overline{g(z)^\nu}$, it is not hard to show that, in general,
the requirement that $\omega$ be real restricts $\nu$ to $+1$ or $-1$,
which yield \eq{ansatz_NS} and \eq{ansatz_alt}, respectively. Therefore,
the alternative {\em ansatz}~(\ref{ansatz_alt}) is given by the choice
$h(\bar{z})=1/\overline{g(z)}$ in (\ref{gensol}).
The solution of Example~3 is recovered from such {\em ansatz}
for $g(z)=z^2$, as a direct calculation shows.

More generally, choosing $g(z)=z^n$ in \eq{ansatz_alt} leads to the solution
$B=\frac{n^2}{\rho^2\sin^2\!n\phi}$, where polar coordinates $z=\rho e^{i\phi}$ were used.
The requirement that $B$ be single-valued restricts $n$ to
$n=\frac{1}{2},1,\frac{3}{2},\ldots$ (note that $B$ is insensitive to
the change $n\to-n$).
In this way, the solution of Example~3 is the $n=2$ case of such family of
two-dimensional singular solutions.
Note that the solution corresponding to a given $n$ is singular
along $2n$ lines starting at the origin and passing through the roots of
unity of order $2n$.

\end{document}